\documentclass[pra,twocolumn,superscriptaddress,showpacs]{revtex4}
\usepackage{amsthm,amssymb,euscript,amscd,stmaryrd}
\usepackage{graphicx}

\def\bbr{{\mathbb R}}
\def\op#1{#1}        
\def\ket#1{| #1 \rangle}
\def\bra#1{\langle #1 |}

\def\lket#1{| #1 \rangle\rangle}
\def\comm#1#2{\llbracket #1, #2\rrbracket}

\def\Tr{\mathop{\rm Tr}}

\def\rmi{{\rm i}}
\def\A{{\cal A}}
\def\B{{\cal B}}

\def\H{{\cal H}}
\def\L{{\cal L}}
\def\e{{\bf e}}
\newif\ifpdflatex\pdflatextrue
\makeatletter\@ifundefined{pdfoutput}{\pdflatexfalse}\makeatother
\def\myincludegraphics[#1]#2#3{%
\ifpdflatex \includegraphics[#1]{#2}
\else       \includegraphics[#1]{#3}
\fi}
\begin{document}
\title{Constraints on relaxation rates for $N$-level quantum systems}
\author{S.~G.~Schirmer} \email{sgs29@cam.ac.uk},
\affiliation{Department of Applied Maths and Theoretical Physics,
         University of Cambridge, Wilberforce Road, Cambridge, CB3 0WA, UK}
\affiliation{Department of Engineering, Division F, Control Group,
         University of Cambridge, Trumpington Street, Cambridge, CB2 1PZ, UK}
\author{A.~I.~Solomon} \email{a.i.solomon@open.ac.uk}
\affiliation{Department of Physics and Astronomy, The Open University,
         Milton Keynes, MK7 6AA, UK}
\affiliation{LPTL, University of Paris VI, 75252 Paris, France}
\date{\today}

\begin{abstract}
We study the constraints imposed on the population and phase relaxation rates
by the physical requirement of completely positive evolution for open $N$-level
systems.  The Lindblad operators that govern the evolution of the system are 
expressed in terms of observable relaxation rates, explicit formulas for the
decoherence rates due to population relaxation are derived, and it is shown that
there are additional, non-trivial constraints on the pure dephasing rates for 
$N>2$.  Explicit, \emph{experimentally testable} inequality constraints for the
decoherence rates are derived for three and four-level systems, and the 
implications of the results are discussed for generic ladder, $\Lambda$ and $V$
systems and transitions between degenerate energy levels.
\end{abstract}
\maketitle

\section{Introduction}
\label{sec:intro}

Understanding the dynamics of open systems is crucial in many areas of physics 
including quantum optics~\cite{97Scully, 00Gardiner}, quantum measurement 
theory~\cite{83Kraus}, quantum state diffusion~\cite{98Percival}, quantum 
chaos~\cite{00Braun}, quantum information processing~\cite{00Nielsen} and
quantum control~\cite{PRA64n012414,PRA65n010101,PRA65n042301}.  Yet, despite
many efforts to shed light on these issues~\cite{76Davies,83Kraus,87Alicki,
00Braun,02Breuer}, many important questions remain.  

For instance, it was recognized early by Kraus~\cite{71Kraus}, Lindblad~%
\cite{75Lindblad,76Lindblad}, and Gorini, Kossakowki and Sudarshan~\cite{76Gorini} 
that the dynamical evolution of an open system must be completely positive~%
\footnote{A map $\Lambda$ acting on the algebra of bounded operators $B(\H)$ 
on a finite-dimensional Hilbert space $\H$ is completely positive if the 
composite map $\Lambda \otimes I_n$ acting on $B(\H)\otimes M(n)$, where 
$M(n)$ is algebra of complex $n \times n$ matrices and $I_n$ the identity 
in $M(n)$, is positive for any $n\ge 0$.}
to ensure that the state of the open system remains physically valid at all 
times.  Unfortunately, if relaxation rates are introduced ad-hoc based on a 
phenomenological description of the system, the resulting equations often do
not satisfy this condition.  For example, the Agarwal/Redfield equations of
motion for a damped harmonic oscillator have been shown to violate complete 
and even simple positivity for certain initial conditions~\cite{JCP107p5236}.
Although such master equations may provide physical solutions in some cases,
serious inconsistencies such as negative or imaginary probabilities, unbounded
solutions and other problems may arise.

For two-level systems the implications of the complete positivity requirement
have been studied extensively in the literature, for example by Gorini 
\textit{et al.} who first showed that there are constraints on the relaxation
rates in the weak coupling limit~\cite{76Gorini,78Gorini}, Lendi who provided
a comprehensive in-depth analysis of the dissipative optical Bloch equations%
~\cite{87Alicki}, and more recently Kimura who extended earlier work by Gorini
to the strong coupling regime~\cite{PRA66n062113}.  Recently, there has also 
been considerable research activity on quantum Markov channels for two-level 
systems, motivated by their importance in quantum computing and communication.  
See, for instance, Ref.~\cite{PRA67n062312} for a comprehensive analysis.
A few simple higher dimensional systems such as a three-level $V$-system with 
decay from two upper levels to a common ground state have also been studied,
for instance by Lendi~\cite{87Alicki}.  

In general, however, ensuring complete positivity of the evolution is often 
neglected for open systems with more than two (or degenerate) energy levels.
For instance, the general expressions (6.A.11) in~\cite{95Mukamel} for the 
relaxation rates ensure complete positivity only for two-level systems, as 
we shall show.  For higher-dimensional systems additional constraints must 
be imposed if complete positivity is to be maintained.  One reason for this
neglect of positivity constraints is that, although the general form of the
admissible generators for quantum dynamical semi-groups is known, it can be
difficult to verify whether a proposed dynamical law for an open system is
consistent with positivity requirements.  The main objective of this paper 
is to address is issue.  

The paper is organized as follows.
Starting with a purely phenomenological description of the interaction of an 
open system with its environment in terms of observable population and phase
relaxation rates --- analogous to the $T_1$ and $T_2$ relaxation times for a
two-level system --- we derive a general form for the dissipation superoperator
in section~\ref{sec:liouv}.  We then explicitly demonstrate with simple examples
in section~\ref{sec:constraints} that the relaxation rates cannot be chosen 
arbitrarily if the evolution of the system is to be physical in the sense that
it satisfies complete positivity.  In particular, we show that the phase 
relaxation rates for $N>2$ are correlated even in the absence of population 
relaxation, i.e., there exist constraints on the phase relaxation rates that
are independent of population decay.  To understand the nature of these 
constraints we express the empirically derived relaxation superoperator in 
Lindblad form (Section~\ref{sec:standard}), and show that it can always be 
decomposed into two parts, one accounting for population relaxation and the 
other for pure phase relaxation processes (Section~\ref{sec:decomp}).  This 
decomposition provides a general formula for the decoherence rates induced by
population decay, which is consistent with physical expectations and evidence,
and \emph{additional} positivity constraints on the decoherence rates resulting
from pure phase relaxation for $N>2$.  

In section~\ref{sec:dephasing} we study the implications of these additional 
constraints in depth for three and four-level systems.  In particular, we use
the abstract positivity constraints to derive explicit inequality constraints
for the observable decoherence rates.  Such explicit constraints are important
from a theoretical and practical perspective because they allow us to make
concrete, empirically verifiable predictions about the decoherence rates and
the dynamics of the system, as we show in section~\ref{sec:examples} for several
common, generic three and four-level systems such as ladder, $\Lambda$, $V$ 
and tripod systems, and transitions between doubly degenerate energy levels.  
Experimental data consistent with positivity constraints would be significant
and validate the chosen model for the open system dynamics. On the other hand, 
if the observed relaxation rates for a system do not satisfy the constraints
required to ensure complete positivity, it would be a strong indication that 
the model used is not sufficient to properly describe the dynamics of the 
system.  This does not necessarily mean that the model is useless; it might 
well be adequate for some purposes, but there will be cases where the model 
makes unphysical predictions and better models, consistent with physical 
constraints, are required.

\section{Quantum Liouville Equation for Dissipative Systems}
\label{sec:liouv}

The state of an $N$-level quantum system is usually represented by a density 
operator $\op{\rho}$ acting on a Hilbert space $\H$.  If the system is closed
then its evolution is given by the quantum Liouville equation
\begin{equation} \label{eq:Liouville}
  \rmi\hbar \frac{d}{dt} \op{\rho}(t) = \comm{\op{H}}{\op{\rho}(t)},
\end{equation}
where $\op{H}$ is the Hamiltonian.  Formally, the dynamics of an open system 
$S$ that is part of a closed supersystem $S+E$ (possibly the entire universe) 
is determined by the Hamiltonian dynamics (\ref{eq:Liouville}) of $S+E$, and 
the state of the subsystem $S$ can be obtained by taking the partial trace of
the entire system's density operator $\op{\rho}_{S+E}$ over the degrees of
freedom of the environment $E$.  Often however, the evolution of the (closed) 
super-system is unknown or too complicated and we are interested only in the 
dynamics of $S$. It is therefore useful to define a density operator $\op{\rho}$
based on the degrees of freedom of $S$, and describe its non-unitary evolution
by amending the quantum Liouville equation to account for the non-Hamiltonian
dynamics resulting from the interaction of $S$ with the environment $E$.

In this paper we restrict our attention to the (common) case where the effect
of the environment $E$ leads to population and phase relaxation (decay and
decoherence, respectively) of the system $S$, and ultimately causes it to
relax to an equilibrium state.   To clearly define what we mean by the terms
population and phase relaxation, note that given an $N$-dimensional quantum
system we can choose a complete orthonormal basis $\{\ket{n}:n=1,2,\ldots,N\}$
for its Hilbert space and expand its density operator with respect to this basis:
\begin{equation} \label{eq:rho1}
  \op{\rho} = \sum_{n=1}^N \left[ \rho_{nn} \ket{n}\bra{n}
            + \sum_{n'>n} \rho_{nn'} \ket{n}\bra{n'} +
                           \rho_{nn'}^* \ket{n'}\bra{n} \right].
\end{equation}
Although we can theoretically choose any Hilbert space basis, physically there
is usually a preferred basis.  Since the interaction with the environment usually
causes the system to relax to an equilibrium state that is a statistical mixture
of its energy eigenstates%
~\footnote{An energy eigenstate of the system is a Hilbert space wavefunction
$\ket{n}$ that satisfies $\op{H}\ket{n}=E_n\ket{n}$, where $\op{H}$ is the
Hamiltonian of the system.},
it is sensible to choose a suitable basis of (energy) eigenstates of the system 
for modelling the relaxation process. In this setting the diagonal elements 
$\op{\rho}_{nn}$ in expansion Eq.~(\ref{eq:rho1}) of $\op{\rho}$ determine the 
populations of the (energy) eigenstates $\ket{n}$, and the off-diagonal elements
$\rho_{nn'}$ ($n\neq n'$) are called coherences, since they distinguish coherent
superpositions of energy eigenstates $\ket{\Psi}= \sum_{n=1}^N c_n\ket{n}$ from 
statistical (incoherent) mixtures of energy eigenstates $\op{\rho}=\sum_{n=1}^N
w_n \ket{n}\bra{n}$.

Population relaxation occurs when the populations of the energy eigenstates change,
typically due to spontaneous emission or absorption of quanta of energy at random 
times.  To account for population relaxation as a result of the interaction with 
an environment we must modify the system's quantum Liouville equation~%
(\ref{eq:Liouville}) to:
\begin{equation}\label{eq:poptrans}
  \dot{\rho}_{nn}(t) = -\frac{\rmi}{\hbar}(\comm{\op{H}}{\op{\rho}(t)})_{nn}
                       - \sum_{k\neq n} \gamma_{kn}\rho_{nn}(t)
                       + \sum_{k\neq n} \gamma_{nk}\rho_{kk}(t),
\end{equation}
where $\op{H}$ represents the Hamiltonian dynamics of $S$, and $\gamma_{kn}$ is
the rate of population relaxation from state $\ket{n}$ to state $\ket{k}$, which
depends on the lifetime of state $\ket{n}$, and in case of multiple decay pathways,
the probability for the particular transition, etc.  The $\gamma_{kn}$ are thus by
definition real and non-negative.  Population relaxation necessarily induces phase
relaxation, and we will later derive explicit expressions for the contribution of
population relaxation to the phase relaxation rates.

In general, phase relaxation occurs when the interaction of the system with the
environment destroys phase correlations between quantum states, and thus converts
coherent superposition states into incoherent mixed states.  Since coherence is
determined by the off-diagonal elements in our expansion of the density operator, 
this effect can be modelled as decay of the off-diagonal elements of $\op{\rho}$:
\begin{equation} \label{eq:dephasing}
 \dot{\rho}_{kn}(t) = -\frac{\rmi}{\hbar}(\comm{\op{H}}{\op{\rho}(t)})_{kn}
                      -\Gamma_{kn} \rho_{kn}(t),
\end{equation}
where $\Gamma_{kn}$ (for $k \neq n$) is the dephasing rate of the transition
$\ket{k} \leftrightarrow \ket{n}$.

Hence, population and phase relaxation change the evolution of the system and
force us to rewrite its quantum Liouville equation as:
\begin{equation} \label{eq:dLE}
  \dot{\rho}(t) = -\frac{\rmi}{\hbar}\comm{\op{H}}{\op{\rho}(t)} + L_D[\op{\rho}(t)],
\end{equation}
where $L_D[\op{\rho}(t)]$ is the dissipation (super-)operator determined by the 
relaxation rates.  It is convenient to note here that the $N \times N$ density 
matrix $\op{\rho}(t)$ can be rewritten as an $N^2$ (column) vector, which we 
denote as $\lket{\rho(t)}$, by stacking its columns.  Since the commutator 
$\comm{\op{H}}{\op{\rho}(t)}$ and the dissipation (super-)operator $L_D[\op{\rho}(t)]$
are linear operators on the set of density matrices, we can write (\ref{eq:dLE}) 
in matrix form:
\begin{equation}
  \frac{d}{dt}\lket{\rho(t)} = \left(-\frac{i}{\hbar}\L_H + \L_D\right)\lket{\rho(t)}
\end{equation}
where $\L_H$ and $\L_D$ are $N^2 \times N^2$ matrices representing the Hamiltonian
and dissipative part of the dynamics, respectively.  Comparison with equations 
(\ref{eq:poptrans}) and (\ref{eq:dephasing}) shows that the non-zero elements of
$\L_D$ are:
\begin{equation}
  \begin{array}{rcll}
  (\L_D)_{(m,n),(m,n)}   &=& -\Gamma_{mn} & \quad  m\neq n  \\
  (\L_D)_{(m,m),(m',m')} &=& +\gamma_{mm} & \quad  m\neq m' \\
  (\L_D)_{(m,m),(m,m)}   &=& -\sum_{{k=1 \atop k\neq m}}^N \gamma_{km}
  \end{array} \label{eq:LD}
\end{equation}
where the index $(m,n)$ should be interpreted as $m+(n-1)N$.  $\Gamma_{mn}=
\Gamma_{nm}$ implies $(\L_D)_{(m,n),(m,n)}=(\L_D)_{(n,m),(n,m)}$.

For a three-level system subject to population and phase relaxation, for instance,
equation (\ref{eq:LD}) gives a dissipation super-operator of the form:
\begin{widetext}
\begin{equation} \label{eq:LD3}
 \L_D = -\left[\begin{array}{*{9}{c}}
 \gamma_{21}+\gamma_{31}&0&0&0&-\gamma_{12}&0&0&0&-\gamma_{13} \\
 0&\Gamma_{12}&0&0&0&0&0&0&0 \\
 0&0&\Gamma_{13}&0&0&0&0&0&0 \\
 0&0&0&\Gamma_{12}&0&0&0&0&0 \\
 -\gamma_{21}&0&0&0&\gamma_{12}+\gamma_{32}&0&0&0&-\gamma_{23} \\
 0&0&0&0&0&\Gamma_{23}&0&0&0 \\
 0&0&0&0&0&0&\Gamma_{13}&0&0 \\
 0&0&0&0&0&0&0&\Gamma_{23}&0 \\
 -\gamma_{31}&0&0&0&-\gamma_{32}&0&0&0&\gamma_{13}+\gamma_{23}
 \end{array}\right]
\end{equation}
\end{widetext}
where $\gamma_{kn}$ and $\Gamma_{kn}$ are the population and phase relaxation
rates, respectively.

\section{Physical Constraints on the Dynamical Evolution}
\label{sec:constraints}

Although (\ref{eq:LD}) gives a general form for the dissipation superoperator of
a system subject to population and phase relaxation, not every superoperator 
$\L_D$ of this form is acceptable on physical grounds since the density operator 
$\op{\rho}(t)$ of the system must remain Hermitian with non-negative eigenvalues
for all $t>0$, and its trace must be conserved~\footnote{%
Trace conservation means that the sum of the populations of all basis states
is preserved, and is equivalent to conservation of probability.  This condition
may be violated, for instance, if the total population of the system is not
conserved, e.g., by atoms being ionized or mapped outside the subspace $S$ or
particles being lost from a trap.  However, this condition is in principle not
restrictive since we can usually amend the Hilbert space $\H_S$ by adding a
subspace $B$ that accounts for population losses from system $S$ so that the
total population of $S+B$ is conserved under the open system evolution resulting
from the interaction of $S+B$ with the environment.}.
It is easy to see that the relaxation parameters in (\ref{eq:LD}) cannot be
chosen arbitrarily if we are to obtain a valid density operator.  For instance, 
it is well known in quantum optics that a two-level atom with decay $\ket{2}
\rightarrow\ket{1}$ at the rate $\gamma_{12}>0$ also experiences dephasing at 
a rate $\Gamma_{12}\ge\frac{1}{2}\gamma_{12}$ since the coherence $\rho_{12}$
must decay with the population of the upper level in order for $\op{\rho}(t)$ 
to remain positive, consistent with the constraints on the relaxation rates for
two-level systems derived in~\cite{76Gorini,PRA66n062113}.

In higher dimensions we also expect population relaxation from state $\ket{n}$ to
$\ket{k}$ at the rate $\gamma_{kn}$ to induce dephasing of this transition at the 
rate $\Gamma_{kn}\ge\frac{1}{2}\gamma_{kn}$.  However, for $N>2$ the situation is
more complicated.  First, a single random decay $\ket{n}\rightarrow \ket{k}$ due 
to spontaneous emission, for instance, may affect other transitions involving the
states $\ket{k}$ or $\ket{n}$.  This is perhaps not too surprising but since it is
a crucial motivation for the following sections, we shall consider two concrete 
examples.

\begin{figure*}
\myincludegraphics[width=3in]{figures/pdf/ev1.pdf}{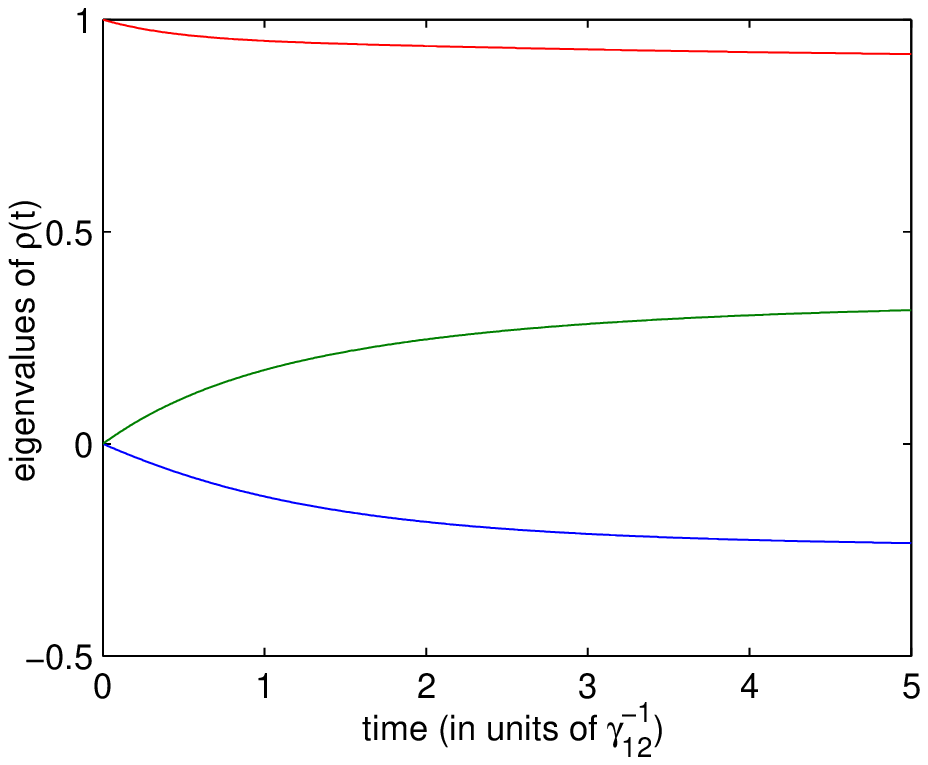} \hfill
\myincludegraphics[width=3in]{figures/pdf/ev2.pdf}{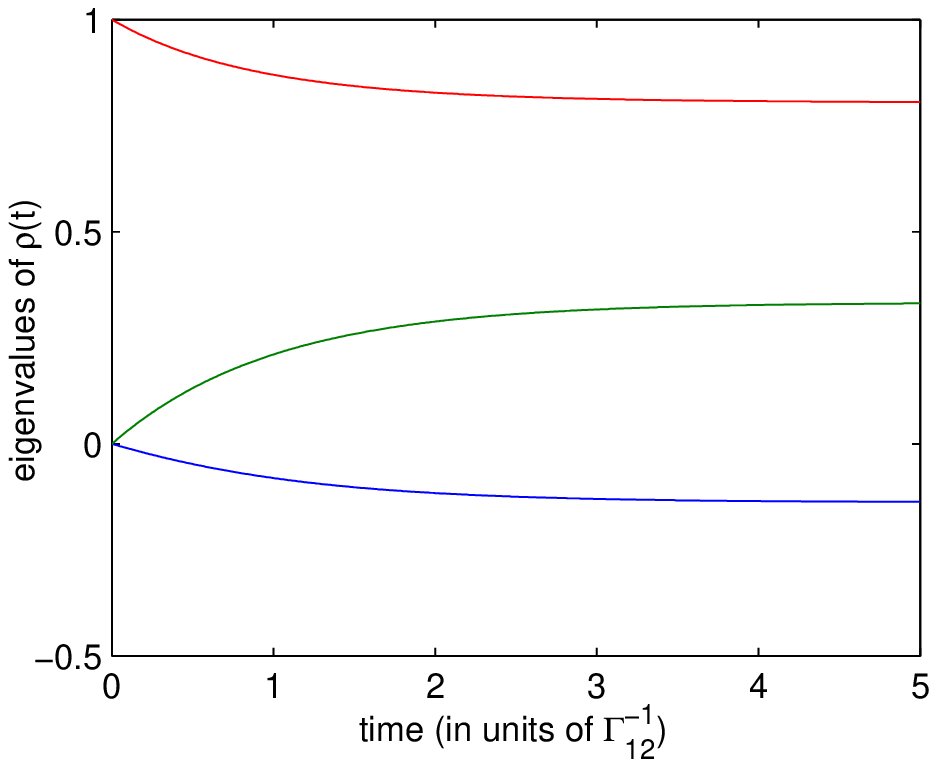}
\caption{Eigenvalues of Eq.~(\ref{eq:false1}), left, and Eq.~(\ref{eq:false2}), 
         right, for $\op{\rho}(0)$ as in Eq.~(\ref{eq:rho0}).}  \label{fig:ev}
\end{figure*}

First, consider a three-level system subject to decay $\ket{2}\rightarrow \ket{1}$ 
at the rate $\gamma_{12}$ but no other relaxation.  Suppose, for instance, we follow 
formula (6.A.11) in \cite{95Mukamel} and set $\Gamma_{12}=\frac{1}{2}\gamma_{12}$ 
and take all other relaxation rates to be zero.  Then, assuming $\op{H}=0$ for 
convenience, the solution of Eq.~(\ref{eq:dLE}) for this dissipation super-operator
leads to the density matrix
\begin{equation} \label{eq:false1}
  \op{\rho}(t) =
  \left( \begin{array}{ccc}
   \rho_{11} + (1-e^{-t\gamma_{12}})\rho_{22} &
   e^{-t\gamma_{12}/2} \rho_{12} & \rho_{13} \\
   e^{-t\gamma_{12}/2} \rho_{21} & e^{-t\gamma_{12}} \rho_{22} & \rho_{23} \\
   \rho_{31} & \rho_{32} & \rho_{33}
  \end{array} \right),
\end{equation}
which in general is not positive for $t>0$.  For example, the superposition state
\begin{equation} \label{eq:rho0}
  \op{\rho}(0)= \ket{\Psi}\bra{\Psi}, \quad \ket{\Psi}=\frac{1}{\sqrt{3}}(1,1,1)^T
\end{equation}
evolves under the action of this dynamical generator to a ``state'' $\op{\rho}(t)$ 
which has a negative eigenvalue (i.e., negative populations) for all $t>0$ as shown 
in Fig.~\ref{fig:ev} and is thus physically unacceptable.

Furthermore, population relaxation is not the only source of constraints on the 
decoherence rates for $N>2$.  A perhaps more surprising observation is that even
if there is no population relaxation at all, i.e., $\gamma_{kn}=0$ for all $k,n$, 
and the system experiences only pure dephasing, we cannot choose the decoherence 
rates $\Gamma_{kn}$ arbitrarily.  For example, setting $\Gamma_{12}\neq 0$ and 
$\Gamma_{23}=\Gamma_{13}=0$ for our three-level system gives
\begin{equation} \label{eq:false2}
  \op{\rho}(t) =
  \left( \begin{array}{ccc}
   \rho_{11} & e^{-\Gamma_{12} t} \rho_{12} & \rho_{13} \\
   e^{-\Gamma_{12} t} \rho_{21} & \rho_{22} & \rho_{23} \\
   \rho_{31} & \rho_{32} & \rho_{33}
  \end{array} \right).
\end{equation}
Choosing $\op{\rho}(0)$ as in Eq.~(\ref{eq:rho0}) we again obtain a density operator
$\op{\rho}(t)$ with negative eigenvalues (see Fig.~\ref{fig:ev}).  This shows that 
there must be additional constraints on the decoherence rates to ensure that the 
state of the system remains physical.

\section{Standard Form of Dissipation Super-Operators}
\label{sec:standard}

Significant progress toward solving the problem of finding dynamical generators
for open systems that ensure complete positivity of the evolution operator, and
hence positivity of the system's density matrix, was made by Gorini, Kossakowski
and Sudarshan \cite{76Gorini} who showed that the generator of a quantum dynamical
semi-group can be expressed in standard form
\begin{eqnarray}
  L[\op{\rho}(t)] &=& -\rmi \comm{\op{H}}{\op{\rho}(t)} \nonumber \\
                  & & \displaystyle + \frac{1}{2}\sum_{k,k'=1}^{N^2-1} a_{kk'}
                   \left( \comm{\op{V}_k \op{\rho}(t)}{\op{V}_{k'}^\dagger}
                  +\comm{\op{V}_k}{\op{\rho}(t)\op{V}_{k'}^\dagger} \right) 
                  \nonumber\\ \label{eq:L1}
\end{eqnarray}
where $\op{H}$ is the generator for the Hamiltonian part of the evolution and
the $\op{V}_k$, $k=1,2,\ldots,N^2-1$, are trace-zero, orthonormal operators
$(\op{V}_k,\op{V}_{k'}):=\Tr(\op{V}_k^\dagger\op{V}_{k'})=\delta_{kk'}$ that
together with $\op{V}_{N^2}=\frac{1}{\sqrt{N}}\op{I}$ form a basis for the
system's Liouville space. Furthermore, the resulting evolution operator is
\emph{completely positive} if and only if the coefficient matrix $a=(a_{kk'})$
is positive.

Noting that a positive matrix $(a_{kk'})$ has real, non-negative eigenvalues
$\gamma_k$ and can be diagonalized by a unitary transformation, we obtain the
second standard representation of dissipative dynamical generator, which was
first derived (independently) by Lindblad~\cite{76Lindblad}:
\begin{eqnarray}  
  L[\op{\rho}(t)] 
  &=& -\rmi \comm{\op{H}}{\op{\rho}(t)} \nonumber\\
  & & + \frac{1}{2}\sum_{k=1}^{N^2-1} 
      \gamma_k \left( \comm{\op{A}_k \op{\rho}(t)}{\op{A}_k^\dagger}+
                      \comm{\op{A}_k}{\op{\rho}(t)\op{A}_k^\dagger} \right). \nonumber\\
    \label{eq:L2}
\end{eqnarray}
Yet, although the general expressions (\ref{eq:L1}) and (\ref{eq:L2}) have been 
known for more than two decades, it is often unknown whether a proposed generator
for the dissipative dynamics for a particular model is completely positive, and 
some common dissipative generators have been shown not to satisfy this condition, 
as in the case of the Agarwal/Redfield equations mentioned earlier. In part this
may be due to the fact that it is often very difficult in practice to express 
phenomenologically derived dissipation generators in either of the two standard 
forms, and hence to verify if a proposed generator satisfies complete positivity.

However, given a matrix representation for the relaxation super-operator of the 
form (\ref{eq:LD}), which was derived from a purely phenomenological model based
on observable decay and decoherence rates, we can express it in standard form 
(\ref{eq:L1}) and transform abstract positivity requirements into concrete, easily
verifiable constraints on the empirically observable relaxation rates.  For this 
purpose we need a basis $\{\op{V}_k\}$ for the Liouville space of the system.  A
canonical choice is to define $N-1$ diagonal matrices
\begin{equation} \label{eq:Vdiag}
  \op{V}_{(m,m)} = \frac{1}{\sqrt{m+m^2}}
                   \left( \sum_{s=1}^m \e_{ss} - m \, \e_{m+1,m+1} \right)
\end{equation}
for $m=1,2,\ldots, N-1$, as well as $N^2-N$ off-diagonal matrices
\begin{equation} \label{eq:Voffdiag}
   \op{V}_{(m,n)} = \e_{mn}, \qquad m\neq n, \; m,n=1,2,\ldots, N.
\end{equation}
where $\e_{mn}$ is an $N\times N$ matrix whose entries are zero except for a one
in $m$th row, $n$th column position.  It is quite easy to verify that the $N^2-1$
operators $\op{V}_{(m,n)}$ thus defined are trace-zero $N\times N$ matrices that
satisfy the orthonormality condition $\Tr(\op{V}_k\op{V}_{k'}^\dagger)=\delta_{kk'}$
for any dimension $N$. 

Having defined the basis operators $\op{V}_k$, we can now compute the generators
\begin{equation} \label{eq:Lkk}
 L_{kk'}[\op{\rho}(t)] 
  = \frac{1}{2}\left(\comm{\op{V}_k \op{\rho}(t)}{\op{V}_{k'}^\dagger}
                    +\comm{\op{V}_k}{\op{\rho}(t)\op{V}_{k'}^\dagger} \right)
\end{equation}
of the dissipation super-operator (\ref{eq:L1}) with respect to this basis, where
$k,k'=1,2,\ldots,N^2-1$. Recalling that $L_{kk'}[\op{\rho}(t)]$ is equivalent to 
$\L_{kk'}\lket{\rho(t)}$, where each $\L_{kk'}$ is an $N^2 \times N^2$ matrix and
$\lket{\rho(t)}$ is an $N^2$-column vector, we note that any (trace-preserving)
dissipation superoperator $\L_D$ can be written as a linear combination of these 
dissipation generators:
\begin{equation} \label{eq:LDexpand}
  \L_D =\sum_{k,k'=1}^{N^2-1} a_{kk'} \L_{kk'}.
\end{equation}
To compute the coefficient matrix $a=(a_{kk'})$ we can rewrite the $N^2\times N^2$
matrices $\L_{kk'}$ and $\L_D$ as column vectors $\vec{\L}_{kk'}$ and $\vec{\L}_D$
of length $N^4$, and $a$ as a column vector $\vec{a}$ of length $(N^2-1)^2$, and 
solve the linear equation $\vec{\L}_D=\A\vec{a}$ where $\A$ is an $N^4\times(N^2-1)^2$
matrix whose columns are the $\vec{\L}_{kk'}$.  This matrix equation has a solution
for any trace-zero Liouville operator since the columns of $\A$ span the space of 
trace-zero Liouville operators of the system.  This procedure allows us in principle
to express any (trace-preserving) dissipation super-operator in standard form 
(\ref{eq:L1}), and to verify whether it generates a completely positive evolution
operator by checking the eigenvalues of the coefficient matrix $(a_{kk'})$.  However,
in practice this is not very efficient, especially for large $N$.  Instead, we would 
like to be able to express the coefficients $a_{kk'}$ directly in terms of observable
relaxation rates.  This is the aim of the following sections.

\section{Decomposition of Relaxation Superoperator}
\label{sec:decomp}

We now use Eq.~(\ref{eq:LDexpand}) to show that the relaxation super-operator of 
any $N$-level system subject to both population and phase relaxation processes 
can be decomposed into a part associated with population relaxation processes and
another accounting for pure decoherence.  To this end, we introduce two types of 
decoherence rates: $\Gamma_{mn}^p$ and $\Gamma_{mn}^d$ for decoherence due to 
population relaxation and pure phase relaxation (dephasing), respectively, and 
require that $\Gamma_{mn}=\Gamma_{mn}^p+\Gamma_{mn}^d$.

If we have population relaxation $\ket{n}\rightarrow\ket{m}$ at the rate 
$\gamma_{mn}\ge 0$ for $m,n=1,2,\ldots,N$ (with $\gamma_{mm}=0$), then setting 
$a_{kk}=\gamma_{mn}$ for $k=m+(n-1)N$ and $a_{kk'}=0$ otherwise in 
(\ref{eq:LDexpand}) leads to a dissipation superoperator
\begin{equation}  \label{eq:LDp}
   \L_D^p = \sum_{m,n=1}^N \gamma_{mn}\L_{(m,n),(m,n)}.
\end{equation}
Inserting Eq.~(\ref{eq:Lkk}) for $\L_{(m,n),(m,n)}$ with $k=m+(n-1)N$ and $\op{V}_k$
as in Eqs~(\ref{eq:Vdiag})--(\ref{eq:Voffdiag}), we obtain
\[
  \begin{array}{lcll}
  (\L_D^p)_{(m,m),(m,m)}   &=& -\sum_{k=1, k\neq m}^N \gamma_{km}, \\
  (\L_D^p)_{(m,m),(m',m')} &=& \gamma_{mm'},      & \quad m\neq m' \\
  (\L_D^p)_{(m,n),(m,n)}   &=& -\frac{1}{2} \sum_{k=1}^N (\gamma_{km}+\gamma_{kn}),
                                                  & \quad m \neq n
\end{array}\]
which agrees with the general form Eq.~(\ref{eq:LD}) of the relaxation super-operator,
yields the correct population relaxation rates, and suggests that the dephasing rates
due to population relaxation are given by:
\begin{equation} \label{eq:Gammap}
 \Gamma_{mn}^p = \frac{1}{2} \sum_{k=1}^N (\gamma_{km}+\gamma_{kn}), \quad m\neq n,
\end{equation}
i.e., that the \emph{decay-induced decoherence} of the transition between states
$\ket{n}$ and $\ket{m}$ is one half of the sum over all decay rates \emph{from}
either of the two states $\ket{m}$ or $\ket{n}$ to any other state.  Finally,
inserting
\begin{equation} \label{eq:Gamma}
 \Gamma_{mn} = \Gamma_{mn}^d+\frac{1}{2} \sum_{k=1}^N (\gamma_{km}+\gamma_{kn}),
   \quad m\neq n
\end{equation}
into Eq.~(\ref{eq:LD}) and solving Eq.~(\ref{eq:LDexpand}) shows that the dissipation
super-operator $\L_D$ of the system decomposes, $\L_D=\L_D^p+\L_D^d$, with $L_D^p$ 
given by Eq.~(\ref{eq:LDp}) and
\begin{equation} \label{eq:LDd}
  \L_D^d = \sum_{m,m'=1}^{N-1} a_{(m,m),(m',m')} \L_{(m,m),(m',m')}.
\end{equation}
Thus, given $\L_D$ and the population relaxation rates $\gamma_{mn}$ of the system, 
we can compute $\L_D^d=\L_D-\L_D^p$ and determine the coefficients $b_{mm'}:=
a_{(m,m),(m',m')}$ in (\ref{eq:LDd}) by rewriting the super-operators $\L_D^d$ and 
$\L_{(m,m),(m',m')}$ as column vectors $\vec{l}$ and $\vec{l}_k$, respectively,
defining a matrix $\B$ whose columns are given by $\vec{l}_k$, and setting $\vec{b}
=\B^{-1}\vec{l}$ where $\B^{-1}$ denotes the pseudo-inverse of $\B$.  However, note
that the matrix $\B$ only has $(N-1)^2$ instead of $(N^2-1)^2$ columns, and we can
eliminate all zero rows.  The resulting coefficient vector $\vec{b}$ can be rearranged
into an $(N-1)\times (N-1)$ coefficient matrix $b=(b_{mm'})$ that depends only on 
the pure dephasing rates $\Gamma_{mn}^d$.  Furthermore, the requirement of positivity
of the coefficient matrix $(a_{kk'})$ in Eq.~(\ref{eq:L1}) now reduces to the (much
simpler) requirement that the $(N-1)\times (N-1)$ matrix $(b_{mm'})$ be positive 
semi-definite.

It is important to note that our formula (\ref{eq:Gammap}) for the contribution of
population relaxation to the overall decoherence rates, obtained solely by imposing
the physical constraint of complete positivity on the evolution of the system, 
agrees with the expressions given, for instance, by Shore \cite{90Shore} for the 
general Bloch equations of $N$-level atoms subject to various dissipative processes,
but our general expression for the dissipation super-operator covers systems subject
to both population decay and pure dephasing processes, and implies the existence of 
non-trivial constraints on the pure dephasing rates of the system for $N>2$.  In the
following sections, we shall analyze these constraints in detail for $N=3$ and $N=4$.

\section{Constraints on the Pure Dephasing Rates}
\label{sec:dephasing}

\subsection{Three-level Systems} 
\label{sec:3level}

Expanding the relaxation super-operator (\ref{eq:LD3}) for a three-level system
with respect to the basis 
\begin{equation} \label{eq:V3}
   \begin{array}{rcl}
   \op{V}_{(1,1)} &=& \frac{1}{\sqrt{2}}(\e_{11}-\e_{22}), \\
   \op{V}_{(2,2)} &=& \frac{1}{\sqrt{6}}(\e_{11}+\e_{22}-2\e_{33})\\
   \op{V}_{(m,n)} &=& \e_{mn}, \quad  m,n=1,2,3, \; m\neq n
   \end{array}
\end{equation}
where $\e_{mn}$ is the $3\times 3$ matrix whose entries are zero except for a 1 
in the $m$th row, $n$th column position --- which corresponds to the canonical 
basis (\ref{eq:Vdiag})--(\ref{eq:Voffdiag}) for $N=3$ --- yields an $8\times 8$ 
coefficient matrix $(a_{kk'})$ whose non-zero entries are $a_{kk}=\gamma_{mn}$ 
for $k=m+3(n-1)$ and $m\neq n$, as well as $a_{(m,m),(m',m')}=b_{mm'}$, where
\[
  \left( \begin{array}{c}
  b_{11} \\
  b_{21} \\
  b_{12} \\
  b_{22}
  \end{array} \right)
  = {\underbrace{\left( \begin{array}{rrrr}
    1           & -\frac{1}{6} \sqrt{3} &  \frac{1}{6} \sqrt{3} & 0 \\
    \frac{1}{4} &  \frac{5}{12}\sqrt{3} &  \frac{1}{12}\sqrt{3} & \frac{3}{4} \\
    1           &  \frac{1}{6} \sqrt{3} & -\frac{1}{6} \sqrt{3} & 0 \\
    \frac{1}{4} & -\frac{5}{12}\sqrt{3} & -\frac{1}{12}\sqrt{3} & \frac{3}{4} \\
    \frac{1}{4} &  \frac{1}{12}\sqrt{3} &  \frac{5}{12}\sqrt{3} & \frac{3}{4} \\
    \frac{1}{4} & -\frac{1}{12}\sqrt{3} & -\frac{5}{12}\sqrt{3} & \frac{3}{4}
  \end{array} \right)}_{\B}}^{-1}
  \left( \begin{array}{c} \Gamma_{12}^d \\
                          \Gamma_{13}^d \\
                          \Gamma_{12}^d \\
                          \Gamma_{23}^d \\
                          \Gamma_{13}^d \\
                          \Gamma_{23}^d
        \end{array} \right).
\]
and the pure dephasing rates $\Gamma_{mn}^d$ are defined by (\ref{eq:Gamma}). 
Noting that the pseudo-inverse of the matrix $\B$ is
\[
  \left( \begin{array}{*{6}{r}}
   \frac{1}{2} & 0 & \frac{1}{2} & 0 & 0 & 0 \\
   \frac{1}{2\sqrt{3}} & \frac{1}{\sqrt{3}} & \frac{1}{2\sqrt{3}} &-\frac{1}{\sqrt{3}} &0& 0\\
   \frac{1}{2\sqrt{3}} & 0&-\frac{1}{2\sqrt{3}} & 0 & \frac{1}{\sqrt{3}}&-\frac{1}{\sqrt{3}}\\
  -\frac{1}{6} & \frac{1}{3} & -\frac{1}{6} & \frac{1}{3} & \frac{1}{3} & \frac{1}{3}
  \end{array} \right)
\]
we obtain
\begin{equation} \label{eq:b3}
  \begin{array}{rcll}
   b_{11} &=& \Gamma_{12}^d \\
   b_{22} &=& (-\Gamma_{12}^d +2\Gamma_{13}^d+2\Gamma_{23}^d)/3\\
   b_{12}=b_{21} &=& (\Gamma_{13}^d-\Gamma_{23}^d)/\sqrt{3}
   \end{array}
\end{equation}
Therefore, the coefficient matrix $(a_{kk'})$ of the relaxation superoperator 
$\L_D$ will be positive semi-definite if and only if $\gamma_{mn}\ge 0$ and 
the real symmetric $2 \times 2$ matrix
\[
  \left( \begin{array}{cc}  b_{11} & b_{12}  \\
                            b_{21} & b_{22}
         \end{array} \right)
\]
has non-negative eigenvalues.  The second condition is equivalent to 
\begin{equation} \label{eq:b3cond}
  b_{11}+b_{22} \ge 0, \quad b_{11} b_{22}\ge b_{12}^2.
\end{equation}
Substituting (\ref{eq:b3}) these conditions become
\begin{equation} \label{eq:Gamma3cond}
   0 \le \Gamma_{12}^d + \Gamma_{13}^d + \Gamma_{23}^d
     \le 2 \sqrt{\Gamma_{12}^d \Gamma_{13}^d
               + \Gamma_{12}^d \Gamma_{23}^d + \Gamma_{13}^d\Gamma_{23}^d}
\end{equation}
This double inequality provides upper and lower bounds for the pure dephasing 
rates of the system.  As expected we can also show that each relaxation rate 
must be non-negative since the second inequality is of the form $[a+(b+c)]^2
\le 4[a(b+c)+bc]$, which can be rewritten as $a^2+(b-c)^2\le 2a(b+c)$, and the
first inequality implies $a+b+c\ge 0$, which shows that $a\ge0$ and $b+c\ge0$,
and by symmetry of $a,b,c$ requires that $\Gamma_{12}^d$, $\Gamma_{13}^d$ and
$\Gamma_{23}^d$ be non-negative.  Hence, we can rewrite (\ref{eq:Gamma3cond})
as follows: 
\begin{equation} \label{eq:Gamma3cond2}
   (\sqrt{\Gamma_b}-\sqrt{\Gamma_c})^2 \le \Gamma_a \le 
   (\sqrt{\Gamma_b}+\sqrt{\Gamma_c})^2 
\end{equation}
where $\{a,b,c\}$ is any permutation of $\{12,13,23\}$.

Note that the choice $\Gamma_{12}^d>0$ and $\Gamma_{13}^d=\Gamma_{23}^d=0$, 
which corresponds to the second example in section~\ref{sec:constraints} if 
there is no population relaxation, clearly violates (\ref{eq:Gamma3cond2}), 
which explains why it results in non-physical evolution.  We also see that 
allowed choices include, for instance, $\Gamma_{12}^d=\Gamma_{23}^d>0$ and 
$\Gamma_{13}^d=0$.

In general, (\ref{eq:Gamma3cond2}) shows that pure dephasing in a three-level
system always affects more than one transition.  Furthermore, if two of the 
pure dephasing rates are equal, say $\Gamma^d$, then the third rate must be 
between $0$ and $4\Gamma^d$.  For instance, consider a triply degenerate 
atomic energy level with basis states $\ket{m=0}$ and $\ket{m=\pm 1}$.  If 
$\Gamma_{-1,0}^d=\Gamma_{0,1}^d$ then $0\le\Gamma_{-1,1}^d\le4\Gamma_{0,1}^d$, 
i.e., the decoherence rate of the transition between the outer states can be 
at most $4\Gamma_{0,1}^d$.  But note in particular that it could be zero even
if the decoherence rate between adjacent states is non-zero.

\subsection{Four-level Systems}
\label{sec:4level}

If we expand the relaxation superoperator $\L_D$ for a four-level system as 
discussed in section~\ref{sec:decomp} with respect to the standard basis 
(\ref{eq:Vdiag})--(\ref{eq:Voffdiag}), we again obtain a coefficient matrix
$(a_{(m,n),(m',n')})$ whose non-zero entries are $a_{(m,n),(m,n)}=\gamma_{mn}$
for $m\neq n$, as well as $a_{(m,m),(m',m')}=b_{mm'}$ with $b_{11}$, $b_{12}$, 
$b_{21}$ and $b_{22}$ as in (\ref{eq:b3}), $b_{31}=b_{13}$, $b_{32}=b_{23}$ and
\begin{equation} \label{eq:b4a}
  \begin{array}{rcl}
  b_{13}&=& \sqrt{6}(-\Gamma_{13}^d+3\Gamma_{14}^d
                     +\Gamma_{23}^d-3\Gamma_{24}^d)/12 \\
  b_{23}&=& \sqrt{2}(-2\Gamma_{12}^d +\Gamma_{13}^d +3\Gamma_{14}^d
                      +\Gamma_{23}^d+3\Gamma_{24}^d-6\Gamma_{34}^d)/12 \\
  b_{33}&=& (-\Gamma_{12}^d-\Gamma_{13}^d+3\Gamma_{14}^d
                      -\Gamma_{23}^d+3\Gamma_{24}^d+3\Gamma_{34}^d)/6
  \end{array}
\end{equation}
Since the reduced coefficient matrix $b=(b_{mm'})$ is a real, symmetric $3 
\times 3$ matrix, \emph{necessary and sufficient} conditions for it to be 
positive semidefinite are~\cite{92Marcus}:
\begin{equation}  \label{eq:b4cond}
   b_{11} \ge 0, \quad
   b_{11} b_{22} \ge b_{12}^2, \quad \det(b) \ge 0.
\end{equation}
The first two of these conditions are equivalent to (\ref{eq:b3cond}). Thus, 
the pure dephasing rates for a four-level system must satisfy (\ref{eq:Gamma3cond}) 
and (\ref{eq:Gamma3cond2}), and the new constraint $\det(b)\ge 0$, or:
\begin{equation} \label{eq:det4}
  b_{11} b_{22} b_{33} + 2 b_{12} b_{13} b_{23}
  \ge  b_{11} b_{23}^2 + b_{22} b_{13}^2 +  b_{33} b_{12}^2.
\end{equation}
Unfortunately, inserting (\ref{eq:b3}) and (\ref{eq:b4a}) into this inequality 
does not yield a nice form for the additional constraint.

We can obtain a more symmetric form of the constraints by choosing a slightly 
different operator basis:
\begin{equation} \label{eq:V4}
   \begin{array}{rcl}
   \op{V}_{(1,1)}' &=& \frac{1}{2}(\e_{11}-\e_{22}+\e_{33}-\e_{44}) \\
   \op{V}_{(2,2)}' &=& \frac{1}{2}(\e_{11}-\e_{22}-\e_{33}+\e_{44}) \\
   \op{V}_{(3,3)}' &=& \frac{1}{2}(\e_{11}+\e_{22}-\e_{33}-\e_{44}) \\
   \op{V}_{(m,n)}' &=& \e_{mn}, \quad m,n=1,2,3,4, \quad m\neq n
   \end{array}
\end{equation}
The $\op{V}_{(m,n)}'$ are trace-zero matrices that differ from the standard 
operator basis only in the choice of the diagonal generators and also form 
an orthonormal basis for the trace-zero Liouville operators of the system. 
However, expanding $\L_D$ with respect to this basis (\ref{eq:V4}) gives a 
more symmetric coefficient matrix $b'$ with non-zero entries:
\begin{equation} \label{eq:b4b}
  \begin{array}{rcl}
   b_{11}' &=& \Gamma_{tot}^d - (\Gamma_{13}^d+\Gamma_{24}^d) \\
   b_{22}' &=& \Gamma_{tot}^d - (\Gamma_{14}^d+\Gamma_{23}^d) \\
   b_{33}' &=& \Gamma_{tot}^d - (\Gamma_{12}^d+\Gamma_{34}^d) \\
   b_{12}' = b_{21}' &=& (\Gamma_{12}^d-\Gamma_{34}^d)/2 \\
   b_{13}' = b_{31}' &=& (\Gamma_{14}^d-\Gamma_{23}^d)/2 \\
   b_{23}' = b_{32}' &=& (\Gamma_{13}^d-\Gamma_{24}^d)/2
  \end{array}
\end{equation}
where $\Gamma_{tot}^d=\frac{1}{2}\sum_{n=2}^4\sum_{m=1}^{n-1}\Gamma_{mn}^d$
is half the sum of all pure dephasing rates.  Since the eigenvalues of the
coefficient matrix are independent of the operator basis, $b'$ has the same
eigenvalues as $b$. Furthermore, \emph{necessary} conditions for $b'$ to
have non-negative eigenvalues are~\footnote{%
To see that these conditions are necessary note that $b_{11}\ge 0$ and $b_{11}
b_{22}\ge b_{12}^2$ implies $b_{22}\ge 0$.  Inserting $b_{12}=b_{23} =b_{13}=0$ 
into (\ref{eq:det4}) yields $b_{11}b_{22}b_{33} \ge 0$ and hence $b_{33}\ge 0$; 
inserting $b_{12}=b_{23}=0$ yields $b_{22}(b_{11}b_{33}-b_{13}^2) \ge 0$, and 
$b_{12}=b_{13}=0$ yields $b_{11}(b_{22}b_{33}-b_{23}^2)\ge 0$. However, these 
conditions are not sufficient since setting $b_{11}=b_{22}=b_{33}=b_{12}=b_{23}
=1$ and $b_{13}=-1$, for instance, satisfies both (\ref{eq:b4cond2a}) and 
(\ref{eq:b4cond2b}) but gives $\det(b)=-4$ and thus violates (\ref{eq:b4cond}).}:
\begin{eqnarray} 
  b_{11} \ge 0, \quad & 
  b_{22} \ge 0, & \quad 
  b_{33} \ge 0,  \label{eq:b4cond2a}\\
  b_{11} b_{22} \ge b_{12}^2,   \quad & 
  b_{11} b_{33} \ge b_{13}^2, & \quad
  b_{22} b_{33} \ge b_{23}^2. \label{eq:b4cond2b}
\end{eqnarray}
Inserting (\ref{eq:b4b}) into (\ref{eq:b4cond2a}) and (\ref{eq:b4cond2b}) yields
\begin{equation} \label{eq:pos2a}
  \begin{array}{rcl}
   \Gamma_{13}^d+\Gamma_{24}^d &\le& \Gamma_{12}^d+\Gamma_{14}^d+\Gamma_{23}^d+\Gamma_{34}^d\\
   \Gamma_{14}^d+\Gamma_{23}^d &\le& \Gamma_{12}^d+\Gamma_{13}^d+\Gamma_{24}^d+\Gamma_{34}^d\\
   \Gamma_{12}^d+\Gamma_{34}^d &\le& \Gamma_{13}^d+\Gamma_{14}^d+\Gamma_{23}^d+\Gamma_{24}^d
\end{array}
\end{equation}
as well as
\begin{equation} \label{eq:pos2b}
\begin{array}{rcl}
 (\Gamma_{14}^d+\Gamma_{23}^d-\Gamma_{13}^d-\Gamma_{24}^d)^2 &\le& 4\Gamma_{12}^d\Gamma_{34}^d\\
 (\Gamma_{12}^d+\Gamma_{34}^d-\Gamma_{13}^d-\Gamma_{24}^d)^2 &\le& 4\Gamma_{14}^d\Gamma_{23}^d\\
 (\Gamma_{12}^d+\Gamma_{34}^d-\Gamma_{14}^d-\Gamma_{23}^d)^2 &\le& 4\Gamma_{13}^d\Gamma_{24}^d
\end{array}
\end{equation}
which can be written simply as
\begin{equation} \label{eq:pos2c}
   |b-c| \le a \le b+c, \quad (b-c)^2 \le 4xy
\end{equation}
if $\{a,b,c\}$ is a permutation of the set $\{\Gamma_{12}^d+\Gamma_{34}^d,
\Gamma_{13}^d+\Gamma_{24}^d,\Gamma_{14}^d+\Gamma_{23}^d\}$ and we let $x$ 
and $y$ be the summands of $a$, e.g., if $a=\Gamma_{12}^d+\Gamma_{34}^d$ 
then $x=\Gamma_{12}^d$ and $y=\Gamma_{34}^d$.  Setting $a=x+y$ shows in 
particular that $0\le x+y$ and $0\le 4xy$ and thus $x,y\ge 0$, from which
we can conclude especially that $\Gamma_{mn}^d\ge0$ for all $m,n$.

In certain cases these constraints can be simplified.  For instance, if the 
pure dephasing rates for transitions between adjacent states are equal, i.e., 
$\Gamma_{12}^d=\Gamma_{23}^d=\Gamma_{34}^d=\Gamma_1^d$, as one might expect, 
for example, for a system consisting of the basis states of a four-fold 
degenerate energy level, and we set $\Gamma_2^d=\frac{1}{2}(\Gamma_{13}^d+ 
\Gamma_{24}^d)$ then (\ref{eq:pos2a}) yields the following bounds on the
decoherence rate $\Gamma_{14}^d$:
\begin{equation}
   \max \{2 \Gamma_2^d-3\Gamma_1^d,\Gamma_1^d-2 \Gamma_2^d,0 \}
   \le \Gamma_{14} \le \Gamma_1^d + 2 \Gamma_2^d
\end{equation}
and combined with the second inequality of (\ref{eq:pos2b}) we obtain $0\le 
\Gamma_2^d \le 4\Gamma_1^d$, and thus $\Gamma_{14}^d \le 9\Gamma_1^d$ and 
$\Gamma_{13}^d,\Gamma_{24}^d \le 8\Gamma_1^d$.
If we further assume that the pure dephasing rates for transitions between
next-to-nearest neighbor states are equal as well, i.e., $\Gamma_{13}^d=
\Gamma_{14}^d=\Gamma_2^d$, then we obtain $\Gamma_{13}^d=\Gamma_{24}^d\le
4\Gamma_1^d$.  Hence, for a system whose pure dephasing rates depend only
on the ``distance'' between the states, the former are bounded above by
\begin{equation} \label{eq:pos4eq}
  \Gamma_{nk}^d \le (n-k)^2 \Gamma_1^d,
\end{equation}
where $\Gamma_1^d=\Gamma_{n,n+1}^d$ is the dephasing rate for transitions
between adjacent sites.

In this special case we can compare the constraints obtained from our 
\emph{necessary} conditions with the \emph{necessary and sufficient} 
conditions (\ref{eq:b4cond}).  Inserting $\Gamma_{nk}^d=\Gamma_{|n-k|}^d$, 
into (\ref{eq:b4b}) yields
\begin{equation}
  \begin{array}{rcl}
   b_{11}' &=& \frac{1}{2}(3\Gamma_1^d -2\Gamma_2^d+\Gamma_3^d) \\
   b_{22}' &=& \frac{1}{2}(\Gamma_1^d+2\Gamma_2^d-\Gamma_3^d) \\
   b_{33}' &=& \frac{1}{2}(-\Gamma_1^d+2\Gamma_2^d-3\Gamma_1^d) \\
   b_{12}' = b_{21}' &=& 0 \\
   b_{13}' = b_{31}' &=& \frac{1}{2} (\Gamma_3^d-\Gamma_1^d) \\
   b_{23}' = b_{32}' &=& 0
  \end{array}
\end{equation}
and the necessary and sufficient conditions (\ref{eq:b4cond}) become:
\begin{equation}
\begin{array}{c}
  \Gamma_3^d \ge 2 \Gamma_2^d-3\Gamma_1^d \\
  \Gamma_3^d \le 2 \Gamma_2^d+\Gamma_1^d \\
  \Gamma_1^d\Gamma_3^d \ge (\Gamma_1^d-\Gamma_2^d)^2
\end{array}
\end{equation}
If $\Gamma_1^d=0$ then $\Gamma_3^d=2\Gamma_2^d=0$. Otherwise, we can multiply
the second inequality by $\Gamma_1^d$ and combine it with the third, which 
leads to $\Gamma_1^d(2\Gamma_2^d+\Gamma_1^d) \ge(\Gamma_1^d-\Gamma_2^d)^2$ and
simplifies to $\Gamma_2^d\le 4\Gamma_1^d$.  Inserting this result in the first
inequality gives $\Gamma_3^d\le 9\Gamma_1^d$, i.e., the necessary conditions 
(\ref{eq:pos4eq}) are also sufficient.

\section{Examples and Discussion}
\label{sec:examples}

We now apply the results of the previous sections to several types of generic
three and four-level atoms.  The objective in each case is to derive a proper
relaxation superoperator, which is consistent with both experimental data and 
positivity constraints, and to discuss the implications of the latter constraints.
Though the emphasis is on atomic systems, the results generally apply to molecular
or solid state systems with similar level structures as well.

\subsection{Generic three-level atoms}
 
Let us first consider the general case of a generic three-level system subject 
to arbitrary population and phase relaxation processes.  Let $\gamma_{mn}$ 
denote the observed rate of population relaxation from state $\ket{n}$ to state
$\ket{m}$ for $m,n=1,2,3$ and $m \neq n$, and let $\Gamma_{12}$, $\Gamma_{23}$ 
and $\Gamma_{13}$ be the observed decoherence rates for the $1 \leftrightarrow 2$, 
$2\leftrightarrow 3$ and $1 \leftrightarrow 3$ transitions, respectively.  Then
the pure dephasing rates of the system according to (\ref{eq:Gamma}) are:
\begin{eqnarray}
\Gamma_{12}^d &=& \Gamma_{12} - (\gamma_{21}+\gamma_{31}+\gamma_{12}+\gamma_{32})/2  
\nonumber \\
\Gamma_{13}^d &=& \Gamma_{13} - (\gamma_{21}+\gamma_{31}+\gamma_{13}+\gamma_{23})/2  
\label{eq:dephasing3}\\
\Gamma_{23}^d &=& \Gamma_{23} - (\gamma_{12}+\gamma_{32}+\gamma_{13}+\gamma_{23})/2 
\nonumber
\end{eqnarray}
These dephasing rates must be non-negative and satisfy the inequality constraints
(\ref{eq:Gamma3cond2}) for the evolution of the system to be completely positive.  
Experimental data for the observed relaxation rates that fails to satisfy these 
conditions should be considered a reason for concern, and might suggest that the
physical system under investigation cannot be adequately modelled as a three-level
system, for instance.  

If the dephasing rates do satisfy the necessary constraints then a physically valid
representation of the relaxation superoperator $L_D$ for the system in terms of the
observed relaxation rates is 
\begin{equation} \label{eq:LD3diag}
  L_D[\op{\rho}(t)]  = \sum_{m\neq n} \gamma_{mn} L_{mn}^p[\op{\rho}(t)]
                       +\delta_1 L_1^d[\op{\rho}(t)]+\delta_2 L_2^d[\op{\rho}(t)],
\end{equation}
where the elementary relaxation terms are
\begin{eqnarray} 
  2L_{mn}^p[\op{\rho}(t)]
  &=& \comm{V_{(m,n)}\op{\rho}(t)}{\op{V}_{(m,n)}^\dagger} 
     +\comm{V_{(m,n)}}{\op{\rho}(t)\op{V}_{(m,n)}^\dagger} \nonumber \\
  2L_m^d[\op{\rho}(t)] 
 &=& \comm{A_m \op{\rho}(t)}{A_m}+\comm{A_m}{\op{\rho}(t) A_m} \nonumber\\
\label{eq:Lmn}
\end{eqnarray}
with $V_{(m,n)}$ as in (\ref{eq:V3}) and the diagonal ``pure dephasing'' 
generators
\begin{eqnarray}
  A_1 = \frac{1}{\sqrt{2x(x-\Delta_1)}}
        \left[\sqrt{3} \Delta_2 V_{(1,1)} + (x-\Delta_1) V_{(2,2)} \right] && \nonumber \\
  A_2 = \frac{-1}{\sqrt{2x(x+\Delta_1)}}
        \left[\sqrt{3} \Delta_2 V_{(1,1)} - (x+\Delta_1) V_{(2,2)} \right], && \nonumber \\ 
\label{eq:A3}
\end{eqnarray}
and the ``effective dephasing'' rates are
\begin{equation}
  \delta_{1/2} = (\Gamma_{12}^d+\Gamma_{13}^d+\Gamma_{23}^d \pm x)/3,
\end{equation}
where $x=\sqrt{\Delta_1^2+3\Delta_2^2}$, $\Delta_1 = 2\Gamma_{12}-\Gamma_{13}-
\Gamma_{23}$ and $\Delta_2 = \Gamma_{13}-\Gamma_{23}$.

\begin{figure}
\begin{center}
\myincludegraphics[width=3.3in]{figures/pdf/3level.pdf}{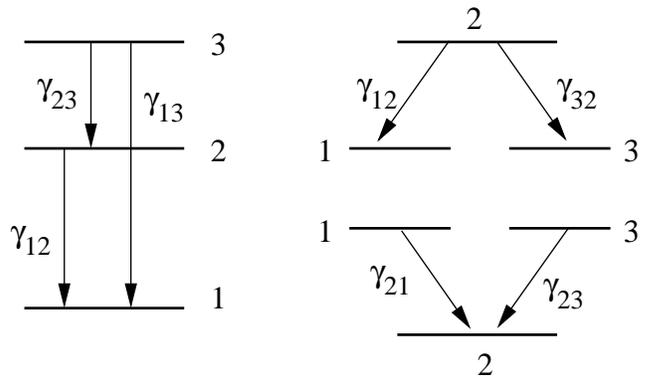}
\caption{Three-state atoms: ladder system (left), $\Lambda$-system (top right)
and V-system (bottom right) with arrows indicating population decay pathways.}
\label{fig:3level}
\end{center}
\end{figure}

\subsubsection{Ladder configurations}
For a three-level atom in a ladder configuration where the main source of population
relaxation is spontaneous emission from the excited states to a stable ground state, 
as shown in Fig.~\ref{fig:3level}, we simply set $\gamma_{21}=\gamma_{31}=\gamma_{32}
=0$ in (\ref{eq:dephasing3}) and (\ref{eq:LD3diag}), respectively, to obtain the 
correct pure dephasing rates
\begin{eqnarray}
\Gamma_{12}^d &=& \Gamma_{12} - \gamma_{12}/2  \nonumber \\
\Gamma_{13}^d &=& \Gamma_{13} - (\gamma_{13}+\gamma_{23})/2  \label{eq:3ladder}\\
\Gamma_{23}^d &=& \Gamma_{23} - (\gamma_{12}+\gamma_{13}+\gamma_{23})/2 \nonumber
\end{eqnarray}
and the corresponding relaxation superoperator $\L_D$. The decay-induced decoherence
rates $\Gamma_{mn}^p$ in this case satisfy the interesting equality $\Gamma_{12}^p+
\Gamma_{13}^p=\Gamma_{23}^p$.  

This is another way of seeing that, if $\gamma_{12}>0$ as in example 1 considered
in section \ref{sec:constraints}, then $\Gamma_{23}$ must be at least $\gamma_{12}/2$
--- recall that we showed explicitly that the naive guess $\Gamma_{12}=\gamma_{12}/2$
and $\Gamma_{13}=\Gamma_{23}=0$ leads to non-physical states with negative eigenvalues.

$\Gamma_{13}=0$, on the other hand, is possible even if $\gamma_{12}>0$ provided that
state $\ket{3}$ is stable.  It is interesting to note, however, that $\Gamma_{13}=0$ 
always implies $\Gamma_{12}=\Gamma_{23}$.  If there is no pure dephasing then this 
is obvious since $\Gamma_{12}^p=\Gamma_{23}^p$, but it is true even if there is pure 
dephasing, since $\Gamma_{13}=0$ implies $\Gamma_{13}^d=0$ and thus the inequality 
constraint (\ref{eq:Gamma3cond2}) for the pure dephasing rates implies $\Gamma_{12}^d
=\Gamma_{23}^d$.

\subsubsection{$\Lambda$ systems}
For a $\Lambda$ system for which only the decay rates $\gamma_{12}$ and $\gamma_{32}$ 
are non-zero as shown in Fig.~\ref{fig:3level}, the pure dephasing rates are:
 \begin{eqnarray}
 \Gamma_{12}^d &=& \Gamma_{12} - (\gamma_{12}+\gamma_{32})/2 \nonumber\\
 \Gamma_{23}^d &=& \Gamma_{23} - (\gamma_{12}+\gamma_{32})/2 \\
 \Gamma_{13}^d &=& \Gamma_{13}. \nonumber
\end{eqnarray}
Moreover, if the lifetime of the excited state is $\gamma^{-1}$ and the system is
symmetric, i.e., $\gamma_{12}=\gamma_{32}=\gamma/2$ and $\Gamma_{12}=\Gamma_{23}$, 
as is often the case, then we have $\Gamma_{12}=\Gamma_{23}=\Gamma^d+\gamma/2$.
If $\Gamma^d=0$ then $\Gamma_{13}^d=0$ due to (\ref{eq:Gamma3cond}).  Otherwise,
setting $\Gamma_{13}^d=\alpha\Gamma^d$ gives $\Delta_1=2(1-\alpha)\Gamma^d$, 
$\Delta_2=-(1-\alpha)\Gamma^d$, $x=2(1-\alpha)\Gamma^d$, and thus the relaxation
superoperator is simply
\begin{eqnarray*}
  L_D[\op{\rho}(t)] 
  &=& (\gamma/2) \left\{L_{12}^p[\op{\rho}(t)]+L_{32}^p[\op{\rho}(t)] \right\} \\
  & & +[(4-\alpha)\Gamma^d/3] L_1^d[\op{\rho}(t)]
      + \alpha\Gamma^d L_2^d[\op{\rho}(t)] 
\end{eqnarray*}
where the diagonal pure dephasing generators (\ref{eq:A3}) are
\begin{eqnarray}
  A_1=A_1^\dagger &=&(-\sqrt{3}V_{(1,1)}+V_{(2,2)})/2 \nonumber \\
  A_2=A_2^\dagger &=&(V_{(1,1)}+\sqrt{3}V_{(2,2)})/2. \label{eq:A3lambda}
\end{eqnarray}
Positivity requires $\gamma\ge0$, $4-\alpha\ge 0$ and $\alpha\ge 0$, or 
$0\le\Gamma_{13}^d\le 4\Gamma^d$, in accordance with (\ref{eq:Gamma3cond2}) and
previous observations.

\subsubsection{$V$ systems}
Similarly, for a $V$ system for which only the decay rates $\gamma_{21}$ and 
$\gamma_{23}$ are non-zero as shown in Fig.~\ref{fig:3level}, the pure dephasing
rates are simply:
\begin{eqnarray}
\Gamma_{12}^d &=& \Gamma_{12}^d - \gamma_{21}/2 \nonumber\\
\Gamma_{23}^d &=& \Gamma_{23}^d - \gamma_{23}/2 \\
\Gamma_{13}^d &=& \Gamma_{13}^d - (\gamma_{21}+\gamma_{23})/2 \nonumber
\end{eqnarray}
If the system is symmetric, i.e., both excited states have the same lifetime, 
$\gamma_{21}=\gamma_{23}=\gamma$, and $\Gamma_{12}=\Gamma_{23}$ then setting
$\Gamma_{13}^d=\alpha\Gamma^d$ leads to the relaxation superoperator
\begin{eqnarray*}
 L_D[\op{\rho}(t)]
 &=& \gamma \left\{L_{21}^p[\op{\rho}(t)] + L_{23}^p[\op{\rho}(t)] \right\} \\
 & & +[(4-\alpha)\Gamma^d/3] L_1^d[\op{\rho}(t)]
     + \alpha\Gamma^d L_2^d[\op{\rho}(t)] 
\end{eqnarray*}
with $L_{mn}[\op{\rho}(t)]$ and $L_1[\op{\rho}(t)]$ as defined in
(\ref{eq:Lmn}) and the generators $A_1$, $A_2$ as in (\ref{eq:A3lambda}).

\subsubsection{Comparison of $\Lambda$ and $V$ systems}
If the excited states have the same lifetime $\gamma^{-1}$ and the pure dephasing
rate $\Gamma^d$ for transitions between the upper and lower states is the same 
for the $\Lambda$ and $V$ configuration, then we have $\Gamma_{12}=\Gamma_{23}=
\frac{1}{2}\gamma+\Gamma^d$ in both cases, i.e., the overall decoherence rate for 
transitions between ground and excited states is the same for both configurations. 
The main difference, as expected, is the decoherence rate of the $1\leftrightarrow
3$ transition, which is $\Gamma_{13}=\Gamma_{13}^d$ for the $\Lambda$-system, and 
$\Gamma_{13}=\gamma+\Gamma_{13}^d$ for the $V$-system.  

Thus, if pure dephasing of the $1 \leftrightarrow 3$ transition is negligible then
it will remain decoherence free for the $\Lambda$ system but not for the $V$ system.
However, if pure dephasing is taken into account then the transition between the 
degenerate ground states of the $\Lambda$ system may not be decoherence free, and
comparison of the decoherence rates for both systems, $\Gamma_{13}^\Lambda=
(\Gamma_{13}^d)^\Lambda$ and $\Gamma_{13}^V = \gamma+(\Gamma_{13}^d)^V$, shows that
$\Gamma_{13}^\Lambda$ could theoretically even be greater than $\Gamma_{13}^V$ if 
the pure dephasing rate of the transition between the degenerate ground states was 
greater than the decay rate $\gamma$ plus the pure dephasing rate $\Gamma_{13}^d$ 
for the $V$ system.

\subsection{Generic four-level atoms}

Again, we will first consider the general case of a generic four-level system 
subject to arbitrary population and phase relaxation processes.  Let $\gamma_{mn}$ 
denote the observed rate of population relaxation from state $\ket{n}$ to state
$\ket{m}$ for $m,n=1,2,3,4$ and $m \neq n$, and $\Gamma_{mn}$ be the observed 
decoherence rates for the $m \leftrightarrow n$, transitions, as usual.  Then
the pure dephasing rates of the system according to (\ref{eq:Gamma}) are:
\begin{eqnarray}
\Gamma_{12}^d &=& \Gamma_{12} - (\gamma_{21}+\gamma_{31}+\gamma_{41}
                                +\gamma_{12}+\gamma_{32}+\gamma_{42})/2 \nonumber \\
\Gamma_{13}^d &=& \Gamma_{13} - (\gamma_{21}+\gamma_{31}+\gamma_{41}
                                +\gamma_{13}+\gamma_{23}+\gamma_{43})/2 \nonumber\\
\Gamma_{14}^d &=& \Gamma_{14} - (\gamma_{21}+\gamma_{31}+\gamma_{41}
                                +\gamma_{14}+\gamma_{24}+\gamma_{34})/2 \nonumber\\
\Gamma_{23}^d &=& \Gamma_{23} - (\gamma_{12}+\gamma_{32}+\gamma_{42}
                                +\gamma_{13}+\gamma_{23}+\gamma_{43})/2 \nonumber\\
\Gamma_{24}^d &=& \Gamma_{24} - (\gamma_{12}+\gamma_{32}+\gamma_{42}
                                +\gamma_{14}+\gamma_{24}+\gamma_{34})/2 \nonumber\\
\Gamma_{34}^d &=& \Gamma_{23} - (\gamma_{13}+\gamma_{23}+\gamma_{43}
                                +\gamma_{14}+\gamma_{24}+\gamma_{34})/2 \nonumber\\
\label{eq:dephasing4}
\end{eqnarray}
These dephasing rates must satisfy the necessary and sufficient conditions 
(\ref{eq:b4cond}) for complete positivity, and in particular the inequality 
constraints (\ref{eq:pos2a}) and (\ref{eq:pos2b}).  If the data for a given
system does not appear to satisfy these conditions then (unless the data is
unreliable) it should be assumed that the system cannot be \emph{properly}
modelled as a four-level system subject to population and phase relaxation.
As mentioned in the introduction, such models are quite common and may still
be adequate for some purposes but can lead to non-physical results such as 
states with negative eigenvalues, etc.

If the dephasing rates do satisfy the necessary constraints then a physically
valid representation of the relaxation superoperator $L_D$ for the system in 
terms of the observed relaxation rates is 
\begin{equation} \label{eq:LD4}
  L_D[\op{\rho}(t)] = \sum_{m\neq n} \gamma_{mn} L_{mn}^p[\op{\rho}(t)]
                     +\sum_{m,n=1}^3 b_{mn} L_{mn}^d[\op{\rho}(t)]
\end{equation}
where the elementary relaxation terms are
\begin{eqnarray} 
  2L_{mn}^p[\op{\rho}(t)]
  &=& \comm{V_{(m,n)}\op{\rho}(t)}{\op{V}_{(m,n)}^\dagger} 
     +\comm{V_{(m,n)}}{\op{\rho}(t)\op{V}_{(m,n)}^\dagger} \nonumber \\
  2L_{mn}^d[\op{\rho}(t)] 
  &=& \comm{V_{(m,m)}'\op{\rho}(t)}{\op{V}_{(n,n)}'} 
     +\comm{V_{(m,m)}'}{\op{\rho}(t)\op{V}_{(n,n)}'} \nonumber \\
\end{eqnarray}
with $V_{(m,n)}'$ as defined in (\ref{eq:V4}), $V_{(m,n)}=V_{(m,n)}'$ for 
$m\neq n$, and coefficients $b_{mn}$ as in (\ref{eq:b4b}).  Note that ---
unlike in the three-level case --- we chose \emph{not} to diagonalize the 
dephasing superoperator since the general expressions are quite complicated
and do not confer a significant computational advantage.  

\subsubsection{Transition between doubly-degenerate levels}
\begin{figure}
\begin{center}\myincludegraphics[width=3in]{figures/pdf/4level.pdf}{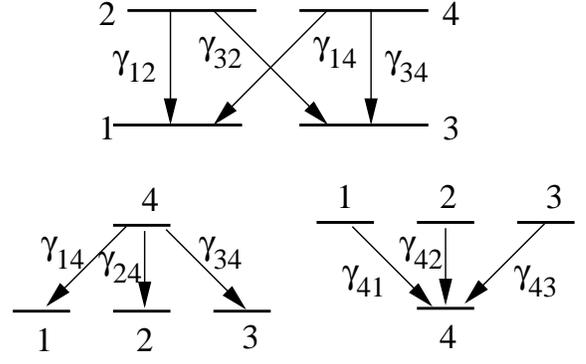}
\caption{Four-state atoms: transition between doubly degenerate energy
levels (top), tripod system (bottom left) and inverted tripod (bottom right)
with arrows indicating population decay pathways.}
\label{fig:4level}
\end{center}
\end{figure}

The results of the last paragraph apply, for instance, to a system consisting of
two doubly degenerate energy levels subject to population relaxation as shown in
Fig.~\ref{fig:4level} and general phase relaxation.  The decay-induced decoherence
rates according to (\ref{eq:Gammap}) are:
\begin{equation} \label{eq:4atomic1}
 \begin{array}{rcl}
 \Gamma_{12}^p=\Gamma_{23}^p &=& (\gamma_{12}+\gamma_{32})/2 \\
 \Gamma_{14}^p=\Gamma_{34}^p &=& (\gamma_{14}+\gamma_{34})/2 \\
 \Gamma_{24}^p               &=& (\gamma_{12}+\gamma_{32}+\gamma_{14}+\gamma_{34})/2\\
 \Gamma_{13}^p               &=& 0.
 \end{array}
\end{equation}
Thus, if all decoherence is the result of population relaxation processes then
the transition between the ground states remains decoherence free.  However, if
there is pure dephasing then (\ref{eq:pos2c}) implies $(b-c)^2\le 4xy$ for $a=
x+y$ with $x=\Gamma_{13}^d$, $y=\Gamma_{24}^d$, $b=\Gamma_{12}^d+\Gamma_{34}^d$
$c=\Gamma_{14}^d+\Gamma_{23}^d$.  Thus, $x=\Gamma_{13}^d=0$ is possible if and
only if $b=c$.  Since $\Gamma_{14}^p+\Gamma_{23}^p=\Gamma_{12}^p+\Gamma_{34}^p$ 
according to (\ref{eq:4atomic1}), this is equivalent to $\Gamma_{12}+\Gamma_{34}
=\Gamma_{14}+\Gamma_{23}$.  Conversely, if $\Gamma_{12}+\Gamma_{34}\neq\Gamma_{14}
+\Gamma_{23}$ then $b\neq c$, and we have $0<|b-c|<x+y$ and $0<(b-c)^2\le 4xy$,
which implies $x>0$, i.e., $\Gamma_{13}^d>0$.

Now suppose both excited states have the same $T_1$-relaxation time, i.e., the 
same spontaneous emission rate $\gamma$, and the relative probabilities for the
possible decay pathways are given by the absolute value of the Clebsch-Gordan 
coefficients of the transition.  Then $\gamma_{12}=\gamma_{34}=\gamma/3$ and 
$\gamma_{32}=\gamma_{14}=2\gamma/3$, and the decay-induced decoherence rates 
are $\Gamma_{13}^p=0$, $\Gamma_{12}^p=\Gamma_{23}^p=\Gamma_{14}^p=\Gamma_{34}^p
=\gamma/2$ and $\Gamma_{24}^p=\gamma$, as one would reasonably expect.  

Furthermore, if the dephasing rates satisfy $\Gamma_{12}^d=\Gamma_{34}^d=:
\Gamma_1^d$ and $\Gamma_{14}^d=\Gamma_{23}^d=:\Gamma_2^d$, as one might expect
due to symmetry for a typical system, then (\ref{eq:pos2c}) implies especially
$(\Gamma_1^d-\Gamma_2^d)^2 \le 4\Gamma_{13}^d\Gamma_{24}^d$.  Thus, if we have
$\Gamma_{13}^d=0$ then we must also have $\Gamma_1^d=\Gamma_2^d$, and conversely, 
if $\Gamma_1^d\neq\Gamma_2^d$ then $\Gamma_{13}^d>0$, i.e., the transition
between the two ground states can remain decoherence free only if $\Gamma_1^d
=\Gamma_2^d$.  This observation may seem trivial but it might be a convenient 
way of ascertaining if the transition between the ground states is decoherence 
free or not by simply measuring the decoherence of the transitions between the
ground and excited states, and the decay rates of the excited states.

Moreover, if $\Gamma_{13}=0$, then we must have $\Gamma_{12}^d=\Gamma_{14}^d=
\Gamma_{23}^d=\Gamma_{34}^d=:\Gamma^d$ according to our previous observations.  
If $\Gamma^d=0$ as well then $\Gamma_{24}^d=0$ due to (\ref{eq:pos2c}) and we
have $L_D^d[\op{\rho}(t)]=0$, i.e., no dephasing takes place.  Otherwise, 
setting $\Gamma_{24}^d=\alpha\Gamma^d$ leads to the simplified relaxation 
superoperator
\begin{eqnarray}
 L_D[\op{\rho}(t)]
 &=& (\gamma/3) \left\{ L_{12}^p[\op{\rho}(t)] + L_{34}^p[\op{\rho}(t)]\right\} \nonumber\\
 & & +(2\gamma/3)\left\{L_{32}^p[\op{\rho}(t)] + L_{14}^p[\op{\rho}(t)] \right\}\nonumber \\
 & & +(4-\alpha)\Gamma^d \left(\comm{A_1 \op{\rho}(t)}{A_1} 
           +\comm{A_1}{\op{\rho}(t) A_1} \right)/4 \nonumber\\
 & & +\alpha\Gamma^d \left(\comm{A_2 \op{\rho}(t)}{A_2}
           +\comm{A_2}{\op{\rho}(t) A_2} \right)/2 \nonumber\\
\end{eqnarray} 
with $A_1=V_{(1,1)}'$, $A_2=(-V_{(2,2)}'+V_{(3,3)}')/\sqrt{2}$ and $L_{mn}^p
[\op{\rho}(t)]$ as defined in (\ref{eq:Lmn}) and $V_{(m,n)}'$ as defined in 
(\ref{eq:V4}).  Again positivity requires $0 \le \alpha \le 4$, i.e., $0 \le
\Gamma_{24}^d\le 4\Gamma^d$, consistent with our previous observations, and
thus provides an upper bound of $4\Gamma^d+\gamma$ on the total decoherence
of the transition between the upper levels.

\subsubsection{Tripod and Inverted Tripod Systems}
Another common type of four-level system is a tripod system, i.e., a transition 
between a triply degenerate ground state and a non-degenerate excited state 
$\ket{4}$.  With population relaxation due to spontaneous emission as indicated
in Fig.~\ref{fig:4level}, the decay-induced decoherence rates according to 
(\ref{eq:Gammap}) are:
\begin{equation} \label{eq:4atomic2}
  \begin{array}{rcl}
  \Gamma_{12}^p = \Gamma_{13}^p = \Gamma_{23}^p &=& 0 \\
  \Gamma_{14}^p = \Gamma_{24}^p = \Gamma_{34}^p 
   &=& (\gamma_{14}+\gamma_{24}+\gamma_{34})/2
  \end{array}
\end{equation}
Assuming that the lifetime of the excited states is $\gamma^{-1}$ and all decay
pathways are equally probable, we obtain $\gamma_{14}=\gamma_{24}=\gamma_{34}=
\gamma/3$ and $\Gamma_{14}^p=\Gamma_{24}^p=\Gamma_{34}^p=\gamma/2$, as well as
\begin{equation}
  L_D^p[\op{\rho}(t)] 
  = (\gamma/3) (\L_{14}^p[\op{\rho}(t)]+\L_{24}^p[\op{\rho}(t)]+\L_{34}^p [\op{\rho}(t)]) 
\end{equation}
with $L_{mn}^p[\op{\rho}(t)]$ as defined in (\ref{eq:Lmn}).

For comparison, the decay-induced decoherence rates (\ref{eq:Gammap}) for an 
inverted tripod, i.e., a transition between a non-degenerate ground state $\ket{4}$
and a three-fold degenerate excited state with population relaxation as indicated
in Fig.~\ref{fig:4level} are:
\begin{equation} \label{eq:4atomic3}
 \begin{array}{rcl}
  \Gamma_{12}^p &=& \frac{1}{2}(\gamma_{41}+\gamma_{42}) \\
  \Gamma_{13}^p &=& \frac{1}{2}(\gamma_{41}+\gamma_{43}) \\
  \Gamma_{23}^p &=& \frac{1}{2}(\gamma_{42}+\gamma_{43}) \\
  \Gamma_{14}^p &=& \frac{1}{2} \gamma_{41} \\
  \Gamma_{24}^p &=& \frac{1}{2} \gamma_{42} \\
  \Gamma_{34}^p &=& \frac{1}{2} \gamma_{43}.
  \end{array}
\end{equation}
and assuming the lifetime of the excited states is $\gamma^{-1}$ we obtain thus
$\gamma_{41}=\gamma_{42}=\gamma_{43}=\gamma$ and $\Gamma_{14}^p=\Gamma_{24}^p
=\Gamma_{34}^p=\gamma/2$, $\Gamma_{12}^p=\Gamma_{23}^p=\Gamma_{13}^p=\gamma$
as well as
\begin{equation}
  L_D^p[\op{\rho}(t)] 
 = \gamma (\L_{41}^p[\op{\rho}(t)]+\L_{42}^p[\op{\rho}(t)]+\L_{43}^p [\op{\rho}(t)])
\end{equation}
with $L_{mn}^p[\op{\rho}(t)]$ as defined in (\ref{eq:Lmn}).

Interestingly, the decay-induced decoherence rates for transitions between the
upper and lower sublevels are the same for both systems.  In absence of pure
dephasing, the only difference between the two cases is that the degenerate 
subspace remains decoherence-free for the tripod system, while the decoherence
rates for the inverted tripod are equal to the spontaneous emission rate $\gamma$
for the upper levels.  This basic situation does not change very much if we add 
pure dephasing since the tripod and inverted tripod system are equivalent as far
as pure dephasing is concerned.  However, there will be additional constraints, 
and we shall in particular study the case $\Gamma_{14}^d=\Gamma_{24}^d=\Gamma_{34}^d
=\Gamma^d$ and $\Gamma_{12}^d=\Gamma_{23}^d=\Gamma_2^d$, which one might expect 
to occur in many systems for reasons of symmetry.  

If $\Gamma^d=0$ then the necessary conditions (\ref{eq:pos2b}) can only be 
satisfied for $\Gamma_2^d=\Gamma_{13}^d=0$, i.e., if there is no pure dephasing
for transitions between the upper and lower sublevels then all pure dephasing 
rates are zero and all decoherence in the system must be due to population 
relaxation.  Hence, $L_D^d[\op{\rho}(t)]=0$.  Otherwise, set $\Gamma_2^d=
\alpha\Gamma^d$ and $\Gamma_{13}^d=\beta\Gamma^d$. Inserting these values into
the coefficient matrix $b'$ [Eq.~(\ref{eq:b4b})] allows us to directly derive 
necessary and sufficient conditions for positivity of this matrix by computing 
its eigenvalues.
\[
  \det(b'/\Gamma^d-\lambda I) =  (\beta-\lambda)(\lambda^2-p\lambda+q)
\]
where $p=3+2\alpha-\beta$ and $q=(4\alpha-\alpha^2-\beta)/2$, shows immediately
that the eigenvalues of $b'$ are $\lambda_1=\Gamma_{13}^d$, and $\lambda_{2/3}=
(p\pm\sqrt{p^2-4q})/2$.  Hence, necessary and sufficient conditions for positive
semi-definiteness of $b'$ are $\Gamma_{13}^d\ge 0$, $p\ge 0$ and $q\ge 0$, and
the dephasing superoperator can be written as:
\begin{eqnarray}
 L_D^d[\op{\rho}(t)]
 &=& \Gamma_{13}^d \left(\comm{A_1 \op{\rho}(t)}{A_1} 
                        +\comm{A_1}{\op{\rho}(t) A_1} \right)/2 \nonumber\\
 & & +\lambda_2 \left(\comm{A_2 \op{\rho}(t)}{A_2} 
                        +\comm{A_2}{\op{\rho}(t) A_2} \right)/2 \nonumber\\
 & & +\lambda_3 \left(\comm{A_3 \op{\rho}(t)}{A_3}
                        +\comm{A_3}{\op{\rho}(t) A_3} \right)/2 \qquad 
\end{eqnarray} 
where we have
\begin{eqnarray*}
 A_1 &=& (V_{(2,2)}'-V_{(3,3)}')/\sqrt{2} \\
 A_2 &=& [4\tilde{\alpha}V_{(1,1)}'+(\tilde{\beta}+x)V_{(2,2)}'
         -(\tilde{\beta}+x)V_{(3,3)}']/(2x) \\ 
 A_3 &=& [4\tilde{\alpha}V_{(1,1)}'+(\tilde{\beta}-x)V_{(2,2)}'
         -(\tilde{\beta}-x)V_{(3,3)}']/(2x) 
\end{eqnarray*}
with $V_{(m,n)}'$ as defined in (\ref{eq:V4}), and $\tilde{\alpha}=\alpha-1$,
$\tilde{\beta}=\beta-1$ and $x=\sqrt{8(\alpha-1)^2+(\beta-1)^2}$.  Furthermore,
$p\ge 0$ implies $\beta\le3+2\alpha$ and $q\ge 0$ is equivalent to $\alpha(4-
\alpha)\ge\beta\ge 0$ and hence implies $0\le\alpha,\beta\le 4$.

\section{Conclusions}
\label{sec:conclusion}

Starting with very basic assumptions we defined a simple yet general relaxation 
superoperator, which should be adequate to describe a wide variety of open 
systems not too strongly coupled to their environment, solely in terms of 
experimentally observable quantities such as the population relaxation and 
decoherence rates of the system, without imposing any restrictions on the 
types of population and phase relaxation that can occur.

The advantage of a relaxation superoperator thus defined is that it can 
describe the observed dissipative dynamics of the system in principle as 
accurately as we can measure the relaxation rates.  Unfortunately, however, 
there are several problems with this approach.  One is that it can lead to  
relaxation superoperators that do not preserve complete or even simple 
positivity, as we have explicitly shown for several examples.  Since any
violation of positivity effectively means negative or even non-real 
probabilities, this is serious problem.

To avoid such problems one must impose constraints on the relaxation rates.
We have analyzed the nature of these basic constraints by expressing our
relaxation superoperator in the standard form for dissipative generators of
quantum dynamical semi-groups derived by Gorini, Kossakowski and Sudarshan.
We have also shown that it is possible to decompose our generic relaxation 
superoperator into two distinct parts associated with population relaxation
and pure dephasing processes, respectively, and that the coefficients of the
Kossakoswki generators for the population relaxation part can be identified 
(usually uniquely) with the observed population relaxation rates, the only 
restriction being the obvious one that the decay rates be non-negative.  Most
importantly, the expressions we obtain for the decoherence rates induced by
population relaxation agree with similar expressions found in the literature.

However, population relaxation is usually not the only source of decoherence.
To account for other sources of decoherence, we have introduced pure dephasing
rates for each transition by subtracting the decoherence induced by population 
relaxation processes from the observed overall decoherence rates.  These pure
dephasing rates define the pure-phase-relaxation superoperator, and we express
the coefficients of the Kossakoswki generators for this part of the relaxation
superoperator explicitly in terms of these pure dephasing rates for three and
four-level systems.  These expressions, unlike the general expressions for the
coefficients of the population relaxation superoperator, are more complicated,
and the requirement of complete positivity results in nontrivial constraints 
on the dephasing rates, which we have analyzed specifically for three- and 
four-level systems, although the same type of analysis can be performed for 
systems of higher dimension.

Finally, we have applied these general results to study their concrete 
implications for several simple but commonly used three- and four-level model
systems such as $\Lambda$ and $V$ systems, tripod and inverted tripod systems
and transitions between doubly degenerate energy levels.  In each case we have
attempted to make \emph{concrete} predictions about inequality constraints and
correlations of the decoherence rates demanded by the requirement of complete 
positivity, which are \emph{experimentally verifiable}.  Such experimental 
tests of the constraints could be useful in various ways. Confirmation of the 
correlations would vindicate the semi-group description of the dynamics.  On 
the other hand, violation of the constraints required by complete positivity
would suggest that our model of the system is not really adequate to capture 
its real dynamics although it may still be useful for certain purposes.

\section{Epilogue: Positive Matrices}
\label{sec:epilogue}

There has been a great deal of mathematical work on the properties of positive
matrices, and papers such as ``Some inequalities for Positive Definite Symmetric
Matrices'' [Siam J. Appl. Math. 19, 679--681 (1970)] by F.~T.~Man, ``The Space 
of Positive Definite Matrices and Gromov's invariant'' [Trans. Am. Math. Soc. 
274, 239 (1982)] by R.~P.~Savage, or ``Positive Definite Matrices and Catalan
numbers'' [Proc. Am. Math. Soc. 79, 177-181 (1980)] by F.~T.~Leighton and 
M.~Newman appear relevant to our problem at first glance.  However, despite the
connection to this work suggested by their titles, they really address rather 
problems.  

F.~T.~Man, for example, studies the problem of comparing positive definite 
symmetric matrices, in particular answering the question under what conditions
$P>Q$, i.e. $P-Q$ positive, implies $P^2>Q^2$ for positive definite symmetric
matrices $P$ and $Q$.  Unfortunately, these results are not applicable to our
problem.  However, they may be relevant for issues such as the comparison of 
density matrices.

R.~P.~Savage considers the space of $n \times n$ positive definite matrices 
$X$ with $\det(X)=1$ under isometries $X \rightarrow A X A^T$ where $A \in 
SL(n,\bbr)$ and shows that it has a collection of simplices preserved by the 
isometries and that the volume of the top-dimensional ones has a uniform
upper bound.  One could perhaps say that the density matrices $\op{\rho}$ of
interest to us are positive, and that the dynamical Lie group of the system
provides isometries of a sort, but our density matrices are positive matrices
of trace one, which actually rules out $\det\op{\rho}=1$, since positivity 
requires the eigenvalues to be non-negative and $\Tr(\op{\rho})=1$ requires 
that the sum of these eigenvalues be one, which implies $\det(\op{\rho})<1$ 
unless $n=1$ and $\rho=1$.  

Similarly, Leighton and Newman show that the number of $n \times n$ integral, 
triple diagonal matrices that are unimodular, positive definite and whose sub
and super diagonal elements are all one, is the Catalan number $\left(2n \over 
n\right)/(n+1)$, which is an interesting mathematical result but not relevant
to our problem since our density matrices, although positive definite, are not
usually tri-diagonal, and even if they were, the elements on the sub and super
diagonal (the coherences) would have to be less than one for normalized density
matrices.

We thank the referee for bringing our attention to the rich mathematical 
literature on the subject of positive matrices.

\acknowledgments
S.G.S thanks A.~Beige, D.~K.~L.~Oi and A.~K.~Ekert (Univ.\ of Cambridge)
and A.~D.~Greentree (Univ.\ of Melbourne) for helpful discussions and 
suggestions, and acknowledges financial support from the Cambridge-MIT 
Institute, Fujitsu and IST grants RESQ (IST-2001-37559) and TOPQIP
(IST-2001-39215).

\bibliography{papers_old,papers,books,sonia}

\begin{thebibliography}{10}

\bibitem{97Scully}
M.~O. Scully and M.~S. Zubairy, {\em Quantum Optics} (Cambridge University
  Press, Cambridge, UK, 1997).

\bibitem{00Gardiner}
C.~W. Gardiner and P. Zoller, {\em Quantum Noise} (Springer Verlag, Heidelberg,
  2000).

\bibitem{83Kraus}
K. Kraus, {\em States, Effects and Operations} (Springer Verlag, Heidelberg,
  1983).

\bibitem{98Percival}
I.~C. Percival, {\em Quantum state diffusion} (Cambridge University Press, New
  York, 1998).

\bibitem{00Braun}
D. Braun, {\em Dissipative quantum chaos and decoherence} (Springer, New York,
  2000).

\bibitem{00Nielsen}
M.~A. Nielsen and I.~L. Chuang, {\em Quantum Computation and Quantum
  Information} (Cambridge University Press, Cambridge, UK, 2000).

\bibitem{PRA64n012414}
A.~M. Childs, I.~L. Chuang, and D.~W. Leung, Phys. Rev. A {\bf 64},  012314
  (2001).

\bibitem{PRA65n010101}
S. Lloyd and L. Viola, Phys. Rev. A {\bf 65},  010101  (2002).

\bibitem{PRA65n042301}
C. Ahn, A.~C. Doherty, and A.~J. Landahl, Phys. Rev. A {\bf 65},  042301
  (2002).

\bibitem{02Breuer}
H.-P. Breuer and F. Petruccione, {\em The theory of open quantum systems}
  (Oxford University Press, Oxford, UK, 2002).

\bibitem{76Davies}
E.~B. Davis, {\em Quantum Theory of Open Systems} (Academic, London, 1976).

\bibitem{87Alicki}
R. Alicki and K. Lendi, {\em Quantum Dynamical Semigroups and Applications}
  (Springer Verlag, Berlin, 1987).

\bibitem{71Kraus}
K. Kraus, Ann. Phys. {\bf 64},  311  (1971).

\bibitem{76Lindblad}
G. Lindblad, Comm. Math. Phys. {\bf 48},  119  (1976).

\bibitem{75Lindblad}
G. Lindblad, Comm. Math. Phys. {\bf 40},  147  (1975).

\bibitem{76Gorini}
V. Gorini, A. Kossakowski, and E.~C.~G. Sudarshan, J. Math. Phys. {\bf 17},
  821  (1976).

\bibitem{JCP107p5236}
D. Kohen, C.~C. Marsden, and D.~J. Tannor, J. Chem. Phys. {\bf 107},  5236
  (1997).

\bibitem{78Gorini}
V. Gorini {\it et~al.}, Rep. Math. Phys. {\bf 13},  149  (1978).

\bibitem{PRA66n062113}
G. Kimura, Phys. Rev. A {\bf 66},  062113  (2002).

\bibitem{PRA67n062312}
S. Daffer, K. Wodkiewicz, and J.~K. McIver, Phys. Rev. A {\bf 67},  062312
  (2003).

\bibitem{95Mukamel}
S. Mukamel, {\em Principles of Nonlinear Optical Spectroscopy} (Oxford
  University Press, New York, 1995).

\bibitem{90Shore}
B.~W. Shore, {\em Theory of coherent atomic excitation} (John Wiley \& Sons,
  New York, 1990).

\bibitem{92Marcus}
M. Marcus and H. Minc, {\em A survey of matrix theory and matrix inequalities}
  (Dover, New York, 1992).

\end{thebibliography}
\bibliographystyle{prsty}
\end{document}